\documentclass[]{aastex631}
\pdfoutput=1
\usepackage{placeins}
\usepackage{cmap}
\usepackage[T1, T2A]{fontenc}
\usepackage[utf8]{inputenc}
\usepackage[english]{babel}

\accepted{to ApJ, 23 August 2021}

\shorttitle{Search for VHE emission from PSR J0218+4232}
\shortauthors{Acciari et al.}
\graphicspath{{./}{figures/}}
\DeclareUnicodeCharacter{2212}{-}
\begin{document}

\title{Search for Very High-Energy Emission from the millisecond pulsar PSR J0218+4232
}

\author[0000-0001-8307-2007]{V.~A.~Acciari}
\affiliation{Inst. de Astrof\'{\i}sica de Canarias, E-38200 La Laguna, and Universidad de La Laguna, Dpto. Astrof\'{\i}sica, E-38206 La Laguna, Tenerife, Spain}
\author[0000-0002-5613-7693]{S.~Ansoldi}
\affiliation{Universit\`a di Udine and INFN Trieste, I-33100 Udine, Italy}
\author[0000-0002-5037-9034]{L.~A.~Antonelli}
\affiliation{National Institute for Astrophysics (INAF), I-00136 Rome, Italy}
\author[0000-0001-9076-9582]{A.~Arbet Engels}
\affiliation{ETH Z\"urich, CH-8093 Z\"urich, Switzerland}
\author{M.~Artero}
\affiliation{Institut de F\'{\i}sica d'Altes Energies (IFAE), The Barcelona Institute of Science and Technology (BIST), E-08193 Bellaterra (Barcelona), Spain}
\author{K.~Asano}
\affiliation{Japanese MAGIC Group: Institute for Cosmic Ray Research (ICRR), The University of Tokyo, Kashiwa, 277-8582 Chiba, Japan}
\author[0000-0002-2311-4460]{D.~Baack}
\affiliation{Technische Universit\"at Dortmund, D-44221 Dortmund, Germany}
\author[0000-0002-1444-5604]{A.~Babi\'c}
\affiliation{Croatian MAGIC Group: University of Zagreb, Faculty of Electrical Engineering and Computing (FER), 10000 Zagreb, Croatia}
\author[0000-0002-1757-5826]{A.~Baquero}
\affiliation{IPARCOS Institute and EMFTEL Department, Universidad Complutense de Madrid, E-28040 Madrid, Spain}
\author[0000-0001-7909-588X]{U.~Barres de Almeida}
\affiliation{Centro Brasileiro de Pesquisas F\'isicas (CBPF), 22290-180 URCA, Rio de Janeiro (RJ), Brazil}
% ...
\collaboration{193}{(MAGIC Collaboration)}

%NoCollaboration Authors
\author[0000-0003-2478-8018]{P.~A.~Caraveo}
\affiliation{INAF-Istituto di Astrofisica Spaziale e Fisica Cosmica Milano, via E. Bassini 15, I-20133 Milano, Italy}
\author{I.~Cognard}
\affiliation{Laboratoire de Physique et Chimie de l'Environnement et de l'Espace -- Universit\'e d'Orl\'eans / CNRS, F-45071 Orl\'eans Cedex 02, France}
\affiliation{Station de radioastronomie de Nan\c{c}ay, Observatoire de Paris, CNRS/INSU, F-18330 Nan\c{c}ay, France}
\author{L.~Guillemot}
\affiliation{Laboratoire de Physique et Chimie de l'Environnement et de l'Espace -- Universit\'e d'Orl\'eans / CNRS, F-45071 Orl\'eans Cedex 02, France}
\affiliation{Station de radioastronomie de Nan\c{c}ay, Observatoire de Paris, CNRS/INSU, F-18330 Nan\c{c}ay, France}
\author[0000-0001-6119-859X]{A.~K.~Harding}
\affiliation{NASA Goddard Space Flight Center, Greenbelt, MD 20771, USA}
\author[0000-0003-1720-9727]{J.~Li}
\affiliation{CAS Key Laboratory for Research in Galaxies and Cosmology, Department of Astronomy, University of Science and Technology of China, Hefei 230026, People's Republic of China}
\affiliation{School of Astronomy and Space Science, University of Science and Technology of China, Hefei 230026, People's Republic of China}
\author[0000-0001-6779-8043]{B.~Limyansky}
\affiliation{Santa Cruz Institute for Particle Physics, Department of Physics, University of California at Santa Cruz, Santa Cruz, CA 95064, USA}
\author{C.~Y.~Ng}
\affiliation{Department of Physics, The University of Hong Kong, Pokfulam Road, Hong Kong, China}
\affiliation{Laboratory for Space Research, The University of Hong Kong, Hong Kong, China}
\author[0000-0002-1522-9065]{D.~F.~Torres}
\affiliation{Institute of Space Sciences (ICE, CSIC), Campus UAB, Carrer de Magrans s/n, E-08193 Barcelona, Spain; and Institut d'Estudis Espacials de Catalunya (IEEC), E-08034 Barcelona, Spain}
\affiliation{Instituci\'o Catalana de Recerca i Estudis Avan\c{c}ats (ICREA), E-08010 Barcelona, Spain}
\author[0000-0001-6566-1246]{P.~M.~Saz~Parkinson}
%\affiliation{Santa Cruz Institute for Particle Physics, Department of Physics and Department of Astronomy and Astrophysics, University of California at Santa Cruz, Santa Cruz, CA 95064, USA}
\affiliation{Santa Cruz Institute for Particle Physics, Department of Physics, University of California at Santa Cruz, Santa Cruz, CA 95064, USA}
\affiliation{Department of Physics, The University of Hong Kong, Pokfulam Road, Hong Kong, China}
\affiliation{Laboratory for Space Research, The University of Hong Kong, Hong Kong, China}
\nocollaboration{9}

\correspondingauthor{Sidika Merve Colak, Brent Limyansky, Pablo Saz Parkinson, Alessia Spolon}
\email{contact.magic@mpp.mpg.de, Brent.Limyansky@ucsc.edu, pablosp@hku.hk}



\begin{abstract}

\noindent PSR J0218+4232 is one of the most energetic millisecond pulsars known and has long been considered as one of the best candidates for very high-energy (VHE; $>$100 GeV) $\gamma$-ray emission. Using 11.5 years of {\it Fermi} Large Area Telescope (LAT) data between 100 MeV and 870 GeV, and $\sim$90 hours of MAGIC observations in the 20 GeV to 20 TeV range, we have searched for the highest energy $\gamma$-ray emission from PSR J0218+4232. Based on the analysis of the LAT data, we find evidence for pulsed emission above 25 GeV, but see no evidence for emission above 100 GeV (VHE) with MAGIC. We present the results of searches for $\gamma$-ray emission, along with theoretical modeling, to interpret the lack of VHE emission. We conclude that, based on the experimental observations and theoretical modeling, it will remain extremely challenging to detect VHE emission from PSR J0218+4232 with the current generation of Imaging Atmospheric Cherenkov Telescopes (IACTs), and maybe even with future ones, such as the Cherenkov Telescope Array (CTA).

\end{abstract}



\section{Introduction} 
\label{introduction}

\noindent PSR J0218+4232 (hereafter J0218) is a millisecond pulsar (MSP) with a period of 2.3 ms in a 2-day orbit with a $\sim0.2 M_\odot$ white dwarf companion \citep{bassa_2003}. It was serendipitously discovered as a steep spectrum, highly polarized source in a low-frequency radio study of an unrelated supernova~\citep{navarro95}. Its broad radio peak, with a large unpulsed component ($\sim50\%$) makes it unusual, suggesting that it may be an {\it aligned rotator}, a pulsar in which the magnetic field is aligned with the axis of rotation. This was later supported by subsequent polarimetry studies~\citep{stairs99}. With a characteristic age $\tau<$0.5 Gyr and a spindown power of 2.4$\times10^{35}$ erg s$^{-1}$, it is one of the youngest and most energetic MSPs known. It has an extremely strong magnetic field at the light cylinder (B$_{LC} \sim 3.2\times10^5$ G), only slightly weaker than young Crab-like pulsars, but several orders of magnitude weaker at the neutron star surface~\citep{saito97}. Like the Crab Pulsar, J0218 also displays giant pulses~\citep{joshi04,knight06}, something very uncommon among MSPs. Its distance, previously estimated to be greater than $6$ kpc, potentially making it the most luminous MSP known~\citep{du14}, has since been revised downwards to $\sim$3 kpc~\citep{Verbiest2014}. Table~\ref{t:timing} provides a summary of the key properties of J0218.

\noindent J0218 was first detected as a steady source of X-rays and $\gamma$-rays using \emph{ROSAT} and EGRET, respectively~\citep{verbunt96}. It was later shown to display non-thermal pulsed X-ray emission~\citep{Kuiper1998}. Like in radio, J0218 has a large unpulsed component in X-rays. It has been detected by most of the X-ray missions, including \emph{BeppoSAX}~\citep{mineo00}, \emph{Chandra}~\citep{kuiper02}, \emph{XMM-Newton}~\citep{webb03}, \emph{RXTE}~\citep{Kuiper04}, and most recently {\it NICER}~\citep{Deneva2019}. Non-thermal hard X-ray emission was detected with {\it NuSTAR} up to $\sim$70 keV~\citep{Gotthelf2017}, with no evidence for a spectral break above these energies, although it must turn over somewhere between 100 keV and 100 MeV, to be consistent with the $\gamma$-ray emission detected at GeV energies~\citep{Gotthelf2017}.

\begin{figure}[ht!]
       \centering
                \includegraphics[width=7in]{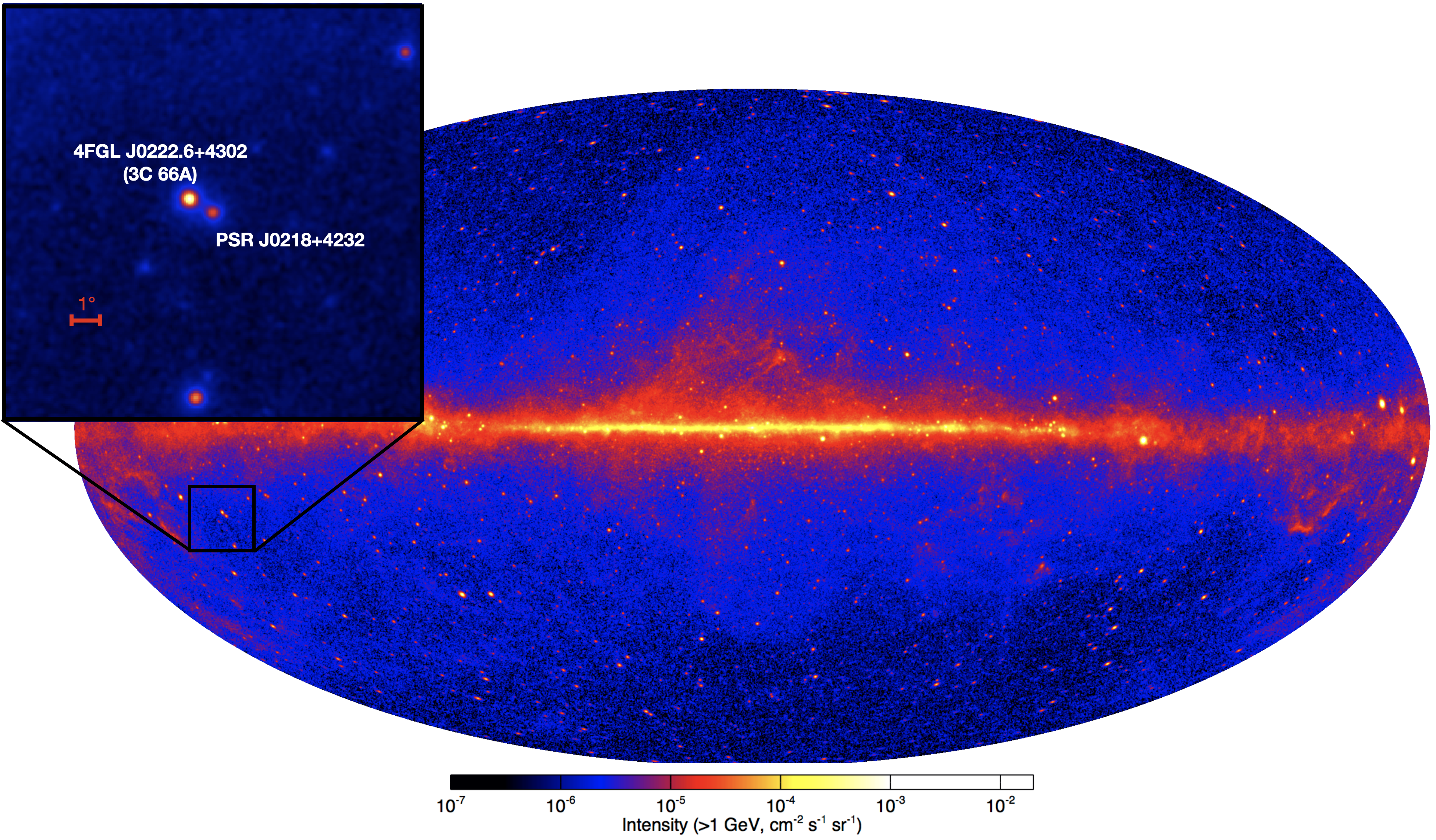}
\caption{LAT All-sky map showing the location of PSR J0218+4232 (aka 4FGL J0218.1+4232). The background map shows the 12-year (August 4, 2008 - August 4, 2020) all-sky intensity map, generated using $\gamma$-ray data above 1 GeV, in Galactic coordinates (Credit: Seth Digel). The square inset region shows the 15$\arcdeg \times 15\arcdeg$ counts map centered on PSR J0218+4232, generated with the 11.5 years of data analyzed in this paper (from 2008 August 4 to 2020 February 10), using all events above 1 GeV. Note the bright $\gamma$-ray blazar 3C 66A (aka 4FGL J0222.6+4302) located less than one degree away from PSR J0218+4232.}  
\label{fig1:LATcountsmap}
\end{figure}

\noindent The $\gamma$-ray emission of J0218 has often been confused with that of the blazar 3C 66A because of their close proximity (0.97$\arcdeg$ separation, see Figure~\ref{fig1:LATcountsmap}), the poor angular resolution of many $\gamma$-ray instruments, and the varying intensity of the blazar. The Second EGRET (2EG) Catalog~\citep{2EG} contained a source (2EG J0220+4228) that was more often associated with 3C 66A. In the third EGRET (3EG) catalog~\citep{3EG}, the source 3EG J0222+4253 was identified with 3C 66A based on its $>$1 GeV emission, however its low-energy flux (100-300 MeV) was found to be dominated by the pulsar, rather than the blazar, leading to the conclusion that both were likely contributing to the EGRET source~\citep{Kuiper00,Guillemot07}. Indeed, \cite{Kuiper00} reported marginal evidence ($\sim3.5\sigma$) for the detection of pulsed $\gamma$-ray emission from J0218, making it potentially the first MSP detected at these energies.

\noindent With the launch of the Large Area Telescope~\citep[LAT,][]{LATinstrument} on board the {\it Fermi} satellite, the picture became significantly clearer. Next to 3C 66A, a strong source ($>19\sigma$), 1FGL J0218.1+4232, was reported in the First \emph{Fermi} LAT Catalog, based on 11 months of data~\citep{1FGL}, and a firm detection of GeV $\gamma$-ray pulsations confirmed the identification with J0218 ~\citep{Abdo2009,1PC}. The LAT detection, however, showed a broad single-peaked $\gamma$-ray light curve, bearing little similarity to the EGRET one reported in \cite{Kuiper00}, or indeed the double-peaked X-ray pulse profile~\citep{kuiper02,webb03}. Despite the fact that the First LAT Catalog of sources above 10 GeV~\citep[][1FHL]{1FHL} contained no source associated with J0218, this pulsar was identified as having hints of pulsed emission above 10 GeV~\citep{1FHL}, and the Third Catalog of Hard {\it Fermi}-LAT Sources~\citep[][3FHL]{3FHL} contained a source associated with J0218 (3FHL J0218.3+4230), which was shown to have $>$10 GeV (and even $>$25 GeV) pulsations~\citep{sazparkinson17}, despite the limited statistics due to the small effective area of the LAT at such high energies, compared to ground-based $\gamma$-ray telescopes\footnote{Note that the sensitivity of ground-based $\gamma$-ray telescopes like MAGIC depends not only on their large effective areas, but also on their ability to reject the cosmic-ray background. Given the challenges of performing background rejection with MAGIC in the 10--100 GeV range, it is perhaps not surprising that MAGIC is less sensitive than $Fermi$-LAT at these energies, despite its much larger effective area.}. These preliminary LAT results provided a strong motivation for observing this pulsar at even higher energies, using ground-based $\gamma$-ray telescopes.

\noindent At very high energies (VHE $>$100 GeV), J0218 has been a target for the Major Atmospheric Gamma Imaging Cherenkov (MAGIC) telescope, starting in 2004, when it was observed for 13 hours during the commissioning phase (single-telescope MAGIC-1 observations), in large part due to it being in the same field of view as 3C 66A~\citep{deona05}. It was subsequently observed for 20 hours, between October 2006 and January 2007, yielding no significant detection, and a 3$\sigma$ flux upper limit of $< 9.4\times10^{-12}$ cm$^{-2}$ s$^{-1}$, above 140 GeV~\citep{anderhub10}. Since then, the performance of the MAGIC telescopes has significantly improved~\citep{Aleksic2016a}.

\noindent In this paper, we report results from an analysis of 11.5 years of {\it Fermi}-LAT data, together with $\sim$90 hours of data from new MAGIC stereoscopic observations of J0218, collected from November 2018 to November 2019, using the low-energy threshold Sum-Trigger-II system \citep{Dazzi2021}. The structure of the paper is as follows: Section~\ref{LATanalysis} describes the {\it Fermi}-LAT data analysis. Section~\ref{MAGIC_analysis} describes our MAGIC observations and analysis. Section~\ref{results} presents our results, both based on LAT data (Section~\ref{LATresults}) and the MAGIC observations (Section~\ref{MAGICresults}). Section~\ref{theory} discusses our theoretical modeling. Finally, in Section \ref{conclusion} we discuss the main conclusions of our work.

\section{Data Analysis}
\subsection{{\it Fermi}-LAT Data Analysis} \label{LATanalysis}

\noindent For the {\it Fermi}-LAT analysis, we used 11.5 years of Pass~8 data~\citep{Atwood2013,Bruel2018}(specifically, P8R3\_SOURCE\_V2), from 2008 August 4 (MJD 54682.7) to 2020 February 10 (MJD 58890).\footnote{We considered the possibility of including {\it Calorimeter-Only} (Cal-Only) data~\citep{Takahashi15,Takahashi19} in our analysis. However, after a preliminary look at 8 years of such data, covering the period 2008-2016, using three different Cal-Only event classes, we found no evidence for pulsed emission, most likely due to the large PSF and corresponding large cosmic-ray background level. Thus, we opted to limit our analysis to {\it standard} LAT data.} We used \texttt{Fermipy}~\citep{wood17} to select {\it Source} class (evclass 128), {\it Front} and {\it Back} converting events (evtype 3) with an energy range from 100 MeV to 870 GeV, and from a square region of 15$\arcdeg \times 15\arcdeg$, centered on the position of 4FGL J0218.1+4232 (RA=34.5344$\arcdeg$, DEC=42.5459$\arcdeg$). A maximum Earth zenith angle of 90$\arcdeg$ is imposed, helping to eliminate contamination from the Earth's limb. We further ensured that the selection only included events at times where the LAT was in normal science configuration and taking good data. 

\begin{figure}[ht!]
       \centering
                \includegraphics[width=7in]{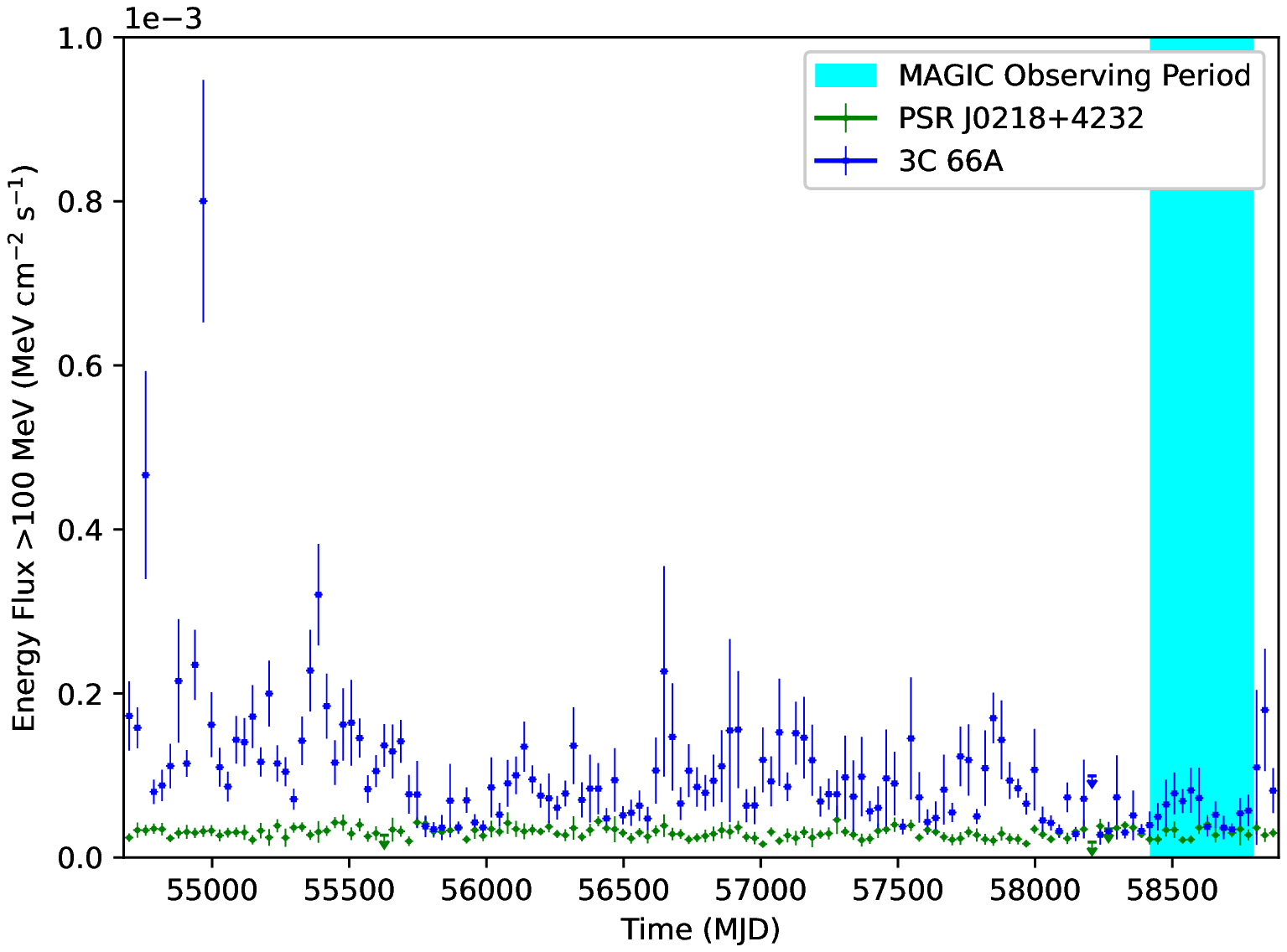}
\caption{Light curve showing PSR J0218+4232 (green circles) and 3C 66A (blue squares). The LAT data ranges from 2008 August 4 (MJD 54682.7) to 2020 February 10 (MJD 58890), and covers the 100 MeV - 870 GeV energy bands. The time period of MAGIC observations (MJD 58424 -- 58791) is shown in cyan. Note the larger variability and $\gamma$-ray flux of 3C 66A.To generate this plot, background sources were fixed to the value in the region model, and the normalizations of 3C 66A and J0218 were allowed to vary. }
\label{fig2:LATlightcurve}
\end{figure}

\begin{figure}[ht!]
       \centering
                \includegraphics[width=7in]{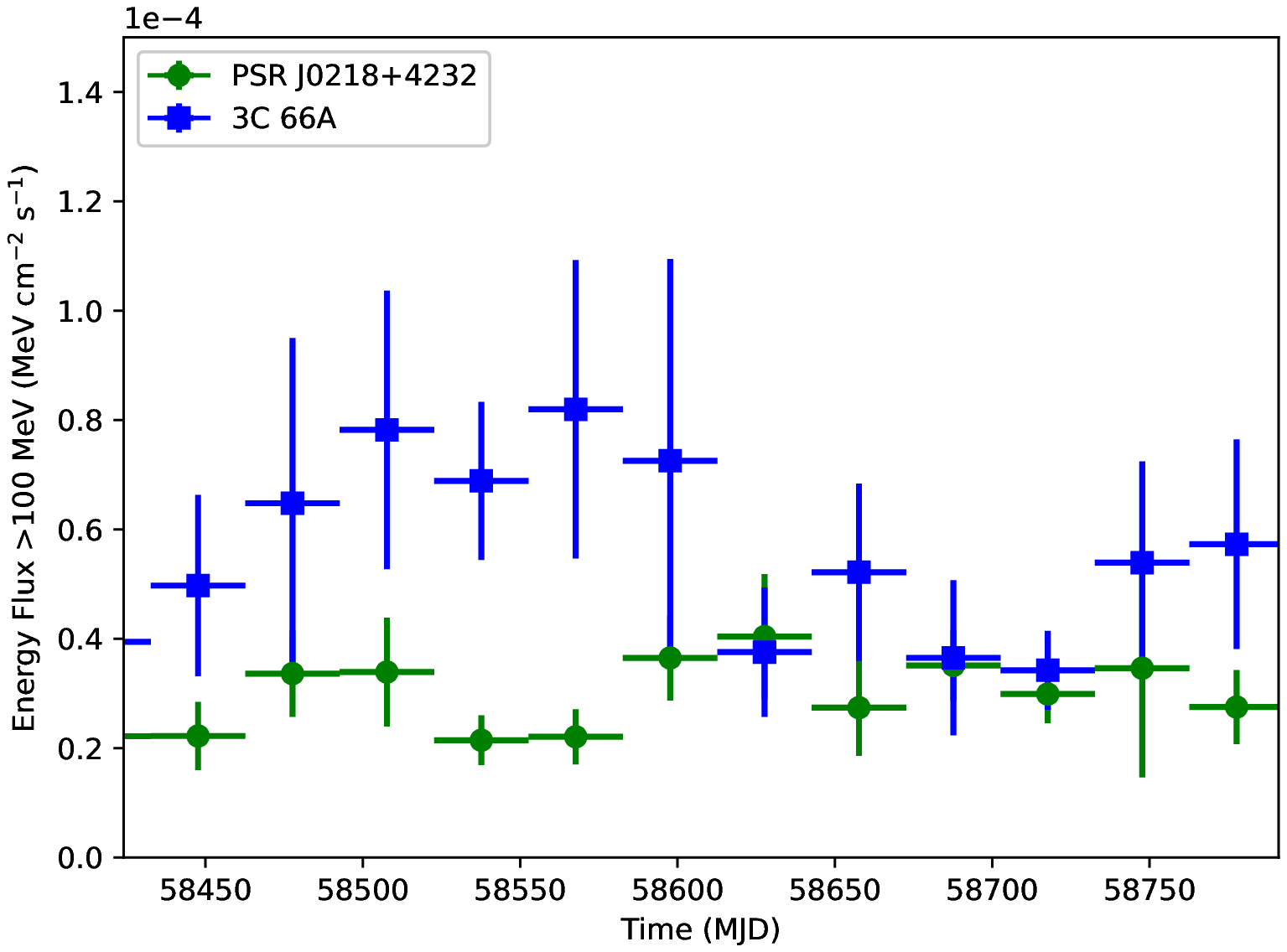}
\caption{\textit{Fermi}-LAT light curve showing PSR J0218+4232 (green circles) and 3C 66A (blue squares), zoomed in on the time period of MAGIC observations (MJD 58424 -- 58791). Note that 3C 66A has a significantly larger flux than J0218 most of the time.
}
\label{fig3:LATlightcurve_zoom}
\end{figure}

\noindent Figure~\ref{fig1:LATcountsmap} shows the {\it Fermi}-LAT $>$1 GeV all-sky $\gamma$-ray intensity map, highlighting the region around J0218. As discussed in Section~\ref{introduction}, the blazar 3C 66A, located 0.97$\arcdeg$ away, complicates the analysis of J0218. The bright $\gamma$-ray counterpart of 3C 66A (4FGL J0222.6+4302, see Figure~\ref{fig1:LATcountsmap}) has a $>$ 1 GeV flux of $1.6\times10^{-8}$ cm$^{-2}$ s$^{-1}$  ($>160\sigma$) in 4FGL, compared to $5.2\times10^{-9}$ cm$^{-2}$ s$^{-1}$ ($ 73.5\sigma$) for J0218. Because blazars are typically variable sources, we considered excluding certain {\it flaring} periods of 3C 66A. Unfortunately, the {\it Fermi}-LAT light curves of J0218 and 3C 66A, over the entire time interval (11.5 years) and energy (100 MeV to 870 GeV) range (Figure~\ref{fig2:LATlightcurve}) reveal that such a strategy would not be possible as the blazar has been quite active throughout the entire 11.5-year period of our observations. Figure~\ref{fig3:LATlightcurve_zoom} shows the zoomed-in 1-year period covered by our MAGIC observations, illustrating how 3C 66A is brighter than J0218 at GeV energies. We note that there is a report of quasi-periodic variability in the optical light curve of 3C 66A with a period of $\sim$3 years, although these have not been confirmed in the $\gamma$-ray data~\citep{Otero-Santos20}. 

\noindent A binned likelihood analysis was performed, utilizing spatial bins of 0.1$\arcdeg \times 0.1\arcdeg$ and 8 logarithmically spaced bins per decade of energy. The initial region model was seeded from the 8-year \textit{Fermi}-LAT Fourth Source Catalog~\citep[4FGL,][]{4fgl} by including sources lying within a square area of 40$\arcdeg \times 40\arcdeg$ centered on 4FGL J0218.1+4232. Because the LAT only measures the energy of photons with finite resolution ($<$10\% in the 1--100 GeV range, but $\sim$20\% at 100 MeV, and $\sim$28\% at 30 MeV)\footnote{\url{https://www.slac.stanford.edu/exp/glast/groups/canda/lat_Performance.htm}} a correction for this energy dispersion was enabled for all sources except the isotropic diffuse emission, as recommended by the {\it Fermi} Science Support Center (FSSC)\footnote{This correction occurs during the fitting of the model. One additional energy bin is added beyond the existing upper and lower energy limits, and filled with a number of photons that is extrapolated from the existing model. The number of photons in each energy bin is then scaled by a factor relating to the instrument response functions, in order to account for the possibility photons with sufficiently large energy uncertainty may have a true energy that lies in a neighboring energy bin. See~\url{https://fermi.gsfc.nasa.gov/ssc/data/analysis/documentation/Pass8_edisp_usage.html}}. The fitting of source spectra within the region model was carried out in an iterative process, with sources being removed from the model if they fell below a test statistic (TS) of 10. Here, TS is defined as TS=$-$2ln($\mathcal{L}_0/\mathcal{L}_1$), where $\mathcal{L}_1$ is the probability that a model which includes the source of interest matches the data, and $\mathcal{L}_0$ is the probability that the same model without the source of interest matches the data\footnote{\url{https://fermi.gsfc.nasa.gov/ssc/data/analysis/documentation/Cicerone/Cicero e_Likelihood/Likelihood_overview.html}}. The final iteration of the fit was performed with Minuit~\citep{James1994}. In addition to the spectral parameters of 4FGL J0218.1+4232, the normalization and index of the Galactic diffuse emission (modeled with a power-law spectrum), the normalization of the isotropic diffuse emission, and the normalizations of background sources with a TS of at least 100 were allowed to vary. 

 \begin{table*}
	\caption{Timing ephemeris for PSR~J0218+4232, obtained with the Nan\c{c}ay radio telescope. We used the DE436 Solar System ephemeris, with time units in barycentric dynamic time (TDB) and the ELL1 binary model for low eccentricity orbits, where EPS1 and EPS2 represent the first and second Laplace-Lagrange parameters~\citep{Lange01}. We refer the reader to the \textsc{Tempo2} manual~\citep{Hobbs2006} for the detailed definition of all parameters included in our timing model.}
	\centering
	\begin{tabular}{ll}
    \hline\hline
    \multicolumn{2}{c}{Timing and binary parameters} \\
    \hline
	R.A., $\alpha$ (J2000.0) \dotfill & $02^{\rm h}\,18^{\rm m}\,06.35863(1)^{\rm s}$ \\
	Decl., $\delta$ (J2000.0) \dotfill & $+42\arcdeg32\arcmin17.3722(2)\arcsec$ \\
	Frequency, $F0$ (Hz) \dotfill & $430.46105998103612106(6)$ \\
	1st frequency derivative, $F1$, (Hz s$^{-1}$) \dotfill &
        $-1.434128(1)\times10^{-14}$ \\
        PMRA ($\dot{\alpha}/cos{\delta}, \mathrm{mas}\,yr^{-1}$)   &        5.32(3)  \\
        PMDEC ($\dot{\delta}, \mathrm{mas}\,yr^{-1}$) &        -3.68(6) \\
        PEPOCH (MJD) & 56000 \\
        POSEPOCH (MJD) & 56000 \\ 
        DMEPOCH  (MJD) &       56000 \\
        DM (cm$^{−3}$ pc) &   61.2374(7) \\
        DM1 (cm$^{−3}$ pc s$^{-1}$)     &   -0.0004(2) \\
        BINARY MODEL &         ELL1 \\
        PB (d)        &     2.0288460845(6) \\
        A1 (lt-s)       &     1.9844348(2) \\
        TASC (MJD)       &     49148.5799767(2) \\
        EPS1     & 5.0(2)$\times 10^{-6}$ \\
        EPS2    &  4.9(2)$\times 10^{-6}$  \\  
        START (MJD)   &   53579.2 \\
        FINISH (MJD) & 58960.5 \\
        UNITS   &       TDB (Barycentric Dynamical Time) \\
        EPHEM   &          DE436\\
        \cutinhead{Derived parameters}
	Period, $P$ (ms) \dotfill & $2.32309053$ \\
	1st period derivative, $\dot{P}$ (s s$^{-1}$) \dotfill & $7.739 \times 10^{-20}$\\
	Characteristic age, $\tau_{\rm c}$ (yr) \dotfill & $4.8 \times 10^{8}$ \\
	Spin-down power, $\dot{E}$ (erg s$^{-1}$) \dotfill & $2.4 \times 10^{35}$ \\
	Surface $B$-field strength, $B_{\rm S}$ (G) \dotfill & $4.3 \times 10^{8}$\\
	Light-cylinder $B$-field, $B_{\rm LC}$ (G) \dotfill & $3.1\times10^{5}$ \\
	Distance, $d$ (kpc) \dotfill & $3.15^{+0.85}_{-0.60}$ \\ 
	ON pulse region \dotfill & (0.34--0.98) \\
	OFF pulse region \dotfill & [0,0.34)$\cup$(0.98,1] \\
	\hline
\end{tabular}
\label{t:timing}
\end{table*}
 
\noindent We modeled the J0218 spectrum using a power law with an exponential cutoff~\footnote{See \url{https://fermi.gsfc.nasa.gov/ssc/data/analysis/scitools/source_models.html}}, $ {dN \over dE} = N_0 ({E \over E_0})^{\gamma} \exp (-aE^b)$. We set the index (b) to a fixed ``sub-exponential" value of 2/3, as this source is too faint for it to be determined by maximum likelihood estimation, and 2/3 approximates the values of other, brighter pulsars~\citep{4fgl}.

\noindent After obtaining our best region model, we examined events within $5\arcdeg$ of J0218, and used the ``\texttt{gtsrcprob}'' \texttt{Fermitool} to assign them a probability ({\it weight}) of originating from either J0218 or 3C 66A relative to other sources in the model. Finally, we used \textsc{Tempo2} \citep{Hobbs2006} with the \texttt{fermi} plug-in \citep{Ray11} to assign the pulsar rotational phases $\phi_i$, according to our pulsar ephemeris obtained with the Nan\c{c}ay radio telescope, given in Table~\ref{t:timing}.

\begin{table*}
	\caption{Gamma-ray spectral parameters for the total emission from} PSR~J0218+4232. Photon and energy flux cover the entire 100 MeV~-~870 GeV energy range.
	\centering
	\begin{tabular}{ll}
	\hline\hline
    \multicolumn{1}{l}{Parameter} & \multicolumn{1}{l}{Value} \\
	\hline
    $N_0$ (ph cm$^{-2}$ s$^{-1}$ MeV$^{-1}$)\dotfill & ($2.07 \, \pm \, 0.03) \times 10^{-11}$ \\
    $\gamma$ \dotfill & $-1.76 \, \pm \,   0.01$ \\
    $E_0$ (MeV) \dotfill & $821.6$ (fixed) \\
    $a$\dotfill & $(6.19751 \, \pm \, 0.00007) \times 10^{-3}$ \\
    $b$ \dotfill & $0.6667$ (fixed) \\
    Photon flux (photons cm$^{-2}$ s$^{-1}$) \dotfill & $(7.67 \, \pm \, 0.15) \times 10^{-8}$\\
	Energy flux (MeV cm$^{-2}$ s$^{-1}$) \dotfill & $(3.05 \, \pm \,   0.04) \times 10^{-5}$ \\
\hline
\end{tabular}
\label{t:spectral}
\end{table*}

\noindent We calculated the source spectrum and flux points for the theoretical modeling described in Section~\ref{theory} by utilizing the aforementioned region model. Three energy bins spanning 12.38 - 28.99 GeV were combined in order to produce a flux point instead of an upper limit. To extract the overall spectrum of J0218, the normalization of the isotropic and Galactic diffuse emission components were allowed to vary while Minuit fit the spectral parameters (summarized in Table~\ref{t:spectral}). To generate flux points, the index of the Galactic diffuse emission was also allowed to vary, along with the normalizations of background sources with a TS of at least 500 or which lie within 5$\arcdeg$ of 4FGL J0218.1+4232. The spectrum of 4FGL J0218.1+4232 is replaced by a power law with an index of -2, and Minuit is used to fit the normalization of this modified spectrum within each energy bin. The result is interpreted as either a flux point or an upper limit, depending on the significance with which the power law source was detected.

\subsection{MAGIC Observations and Data Analysis} \label{MAGIC_analysis}

\noindent We used the MAGIC telescopes to search for the VHE emission component of J0218. MAGIC consists of two imaging atmospheric Cherenkov telescopes (IACTs) of 17m diameter located at the Roque de Los Muchachos Observatory in La Palma, Canary Islands, Spain ~\citep{Aleksic2016a}. We collected the data in stereoscopic mode with the Sum-Trigger-II system. This system is designed to improve the performance of the telescopes in the sub-100 GeV energy range \citep{Dazzi2021}.

\noindent We observed the source from 2018 November 2 to 2019 November 4 (MJD 58424 -- 58791) with a zenith angle range  from 13 to 30 degrees, for maximum sensitivity at low-energies. Wobble mode was used for robust flux and background estimation by pointing the telescopes 0.4$\arcdeg$ away from the source ~\citep{Fomin1994}. Weather conditions were monitored simultaneously by measuring the atmospheric transmission with the LIDAR system operating together with the MAGIC Telescopes ~\citep{Fruck2014}. The Cherenkov radiation produced by the sub-100 GeV particle showers is fainter; consequently, they are more affected by the lower atmospheric transmissions. Therefore, we set a strict requirement of excellent atmospheric conditions, a minimum of 0.85 atmospheric transmission at an altitude of 9km. After discarding around 20\% of the total data, 87 hours of good quality data remained.

\noindent The data were analyzed using the Magic Standard Analysis Software, MARS~\citep{Zanin2013}. We applied the Sum-Trigger-II dedicated algorithm for calibration and image cleaning, which enabled us to improve the performance and achieve an energy threshold of 20 GeV. We performed the higher-level analysis following the standard pipeline \citep{Aleksic2016b}. 

\section{Results} \label{results}

\subsection{{\it Fermi}-LAT results} \label{LATresults}

\begin{figure}[ht!]
       \centering \includegraphics[width=5.5in]{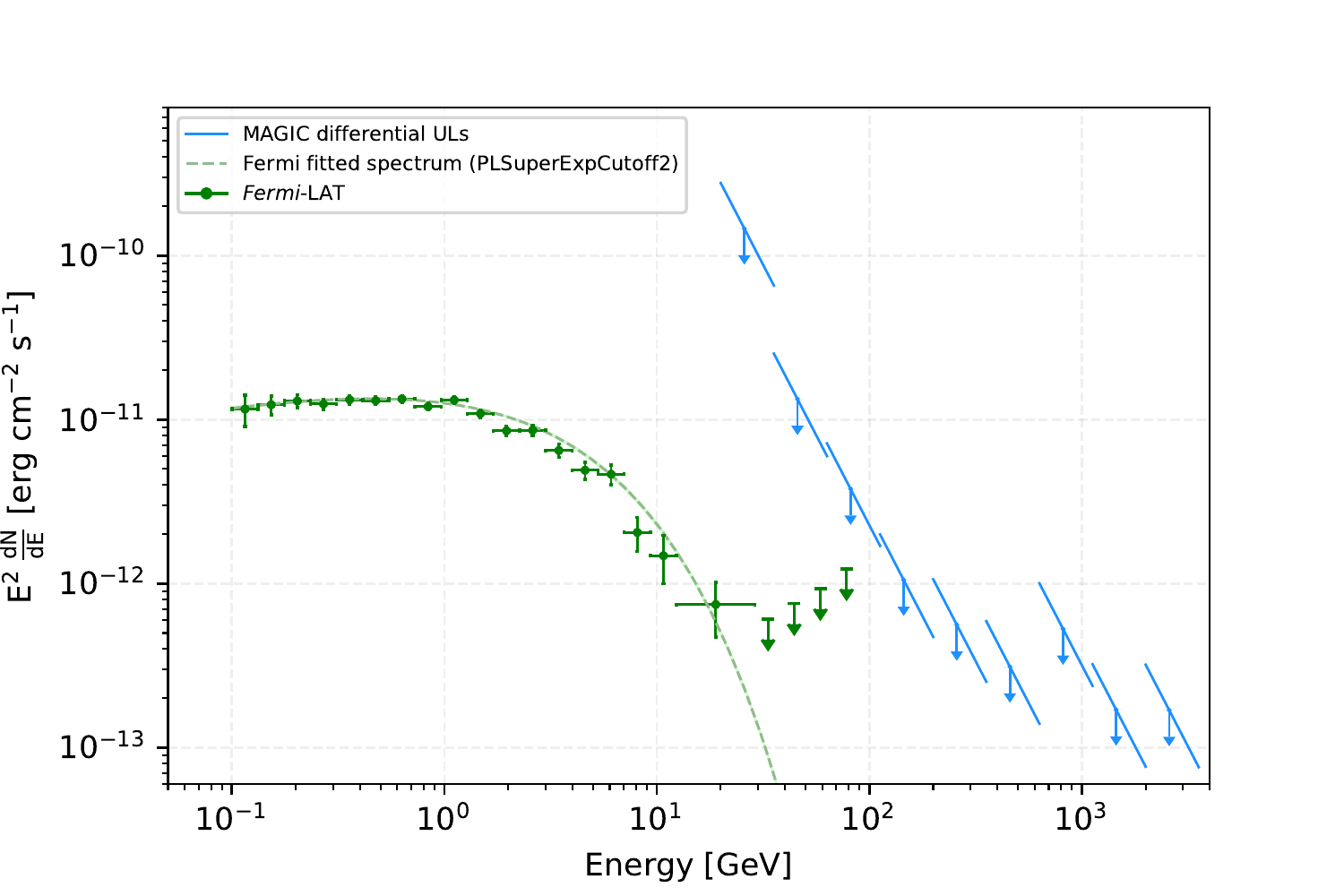}
\caption{Spectrum of the total emission of millisecond pulsar J0218+4232 measured by \textit{Fermi}-LAT (green points and upper limits) and the MAGIC telescopes (light blue upper limits). The green dashes represent the fit of the \textit{Fermi}-LAT data with an exponentially cutoff power-law model. Note that the width of the error region is narrower than the dashes showing the best-fit model. Although included, it is difficult to distinguish in this plot. For the MAGIC analysis we assumed a spectral index $\Gamma=-4.5$ obtained from the spectral index of the power-law fit to the high-energy ($>$10 GeV) part of the \textit{Fermi}-LAT spectrum. }
\label{fig4:LATMAGICSED}
\end{figure}

\noindent Figure~\ref{fig4:LATMAGICSED} shows the LAT spectrum obtained in our analysis. Table~\ref{t:spectral} reports the best-fit spectral parameters and Table~\ref{t:LAT_MAGIC_sed_table} gives the spectral values, including upper limits. Note that the spectrum falls steeply at energies above 10 GeV with only upper limits reported for energies above $\sim$29 GeV. These results are consistent with previous LAT results reported in 4FGL~\citep{4fgl} and 3FHL~\citep{3FHL}, the latter of which reported an index of $\Gamma=-4.5$, when fitting the $>$10 GeV data with a simple power law.

\subsubsection{Search for high-energy pulsation in LAT data} 

\begin{figure}[ht!]
	\includegraphics[width=\columnwidth]{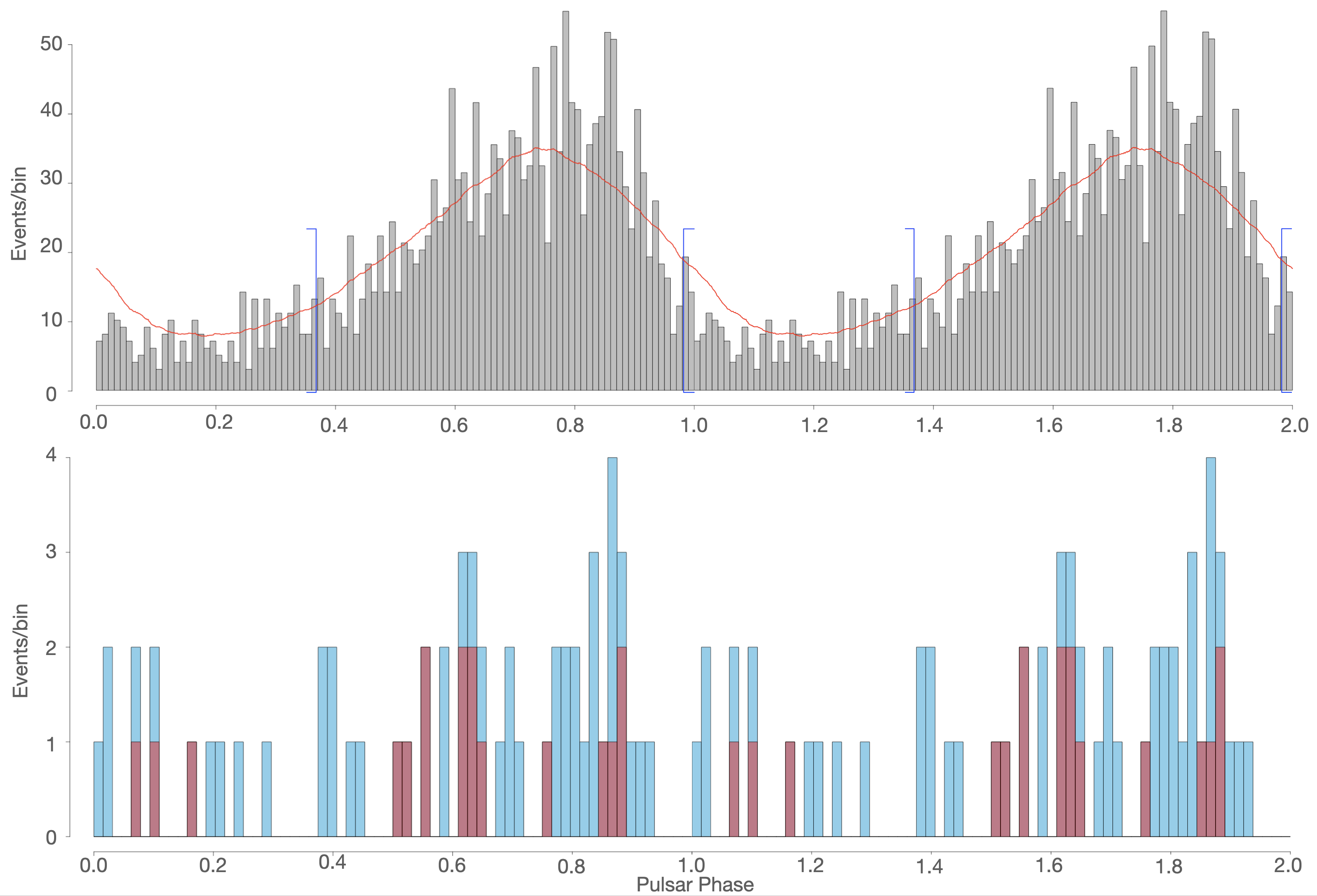}
    \caption{{\bf Top panel -- }Histogram of the 1--10 GeV events for PSR J0218+4232, along with the smooth circular kernel density estimator (red line) fitted to the data, which we define as the {\it low-energy} template in our subsequent searches for pulsed emission above 10 GeV. Two rotation cycles are shown, with 100 bins per cycle. The blue brackets indicate the estimated off-pulse interval, [0--0.34)$\cup$(0.98,1], obtained using SOPIE~\citep{SOPIE}.
    {\bf Bottom panel -- } Search for high-energy pulsations using LAT standard events above 10 GeV. The blue histogram are the 58 events above 10 GeV in energy, while the pink histogram are the 17 events above 25 GeV. Two rotation cycles are shown, with 65 bins per cycle.}
    \label{fig5:Fig_SOPIE}
\end{figure}

\noindent To test for possible pulsed emission above 10 GeV, an analysis analogous to what was carried out in the First \textit{Fermi}-LAT Catalog of $>$10 GeV sources (1FHL) \citep{1FHL} and in \cite{sazparkinson17} was performed. We defined a {\it low-energy} probability density function (PDF), which we refer to as PDF$_{LE}$, based on the best estimate fit of the 1-10 GeV events (see Figure~\ref{fig5:Fig_SOPIE}, top panel). For the high-energy PDF, we considered the family of distributions given by PDF$_{HE}(\phi) = (1 - x) + x \cdot $PDF$_{LE}(\phi)$, with $0 \leq x \leq 1$. We maximized the unbinned likelihood function derived from this PDF, with respect to x, obtaining $\mathcal{L}(\hat{x})$, and comparing it to the null hypothesis, for x=0, that there is no pulsation (i.e. PDF$_{HE}(\phi)$=1). By construction, $\mathcal{L}(0)=1$, so the test statistic (TS=$-$2ln($\mathcal{L}(0)/\mathcal{L}(\hat{x})$) can be simplified to TS = 2 ln$\mathcal{L}(\hat{x})$. We converted the measured TS value into a tail probability (or p-value), by assuming, following Wilks’ theorem~\citep{wilks1938}, that for the null hypothesis the TS follows  a $\chi^2$ distribution with 1 degree of freedom.

\noindent Using the LAT data set described in Section~\ref{LATanalysis}, we first selected events in the 1--10 GeV energy range (with a probability $>$50\% of coming from the pulsar, as obtained via the Fermi Science Tool\footnote{\url{https://fermi.gsfc.nasa.gov/ssc/data/analysis/software/}} {\it gtsrcprob}) and used this histogram to generate a `low-energy template'. We used the non-parametric SOPIE (Sequential Off-Pulse Interval Estimation) R package~\citep{SOPIE}, to obtain a {\it smooth} kernel density estimator of our histogram (which we defined as our ``low-energy template"), and also derived an estimate of the off-pulse interval of the light curve, using the median value of the results obtained from four different goodness-of-fit tests: Kolmogorov–Smirnov, Cram\'{e}r–von Mises, Anderson–Darling, and Rayleigh test statistics. Figure~\ref{fig5:Fig_SOPIE} (top panel) shows our results, including the 1--10 GeV histogram, along with the resulting \textit{low-energy} template, and the estimated off-pulse interval, all calculated using SOPIE. 

\noindent To test for emission at higher energies, we looked at the $>$10 GeV events arriving within the 95\% containment radius of the point-spread function (0.5/0.8 degrees for front/back converting events) and performed a likelihood test to determine whether they are likely to come from a similar distribution function, as represented by the lower energy template. We set a threshold p-value of 0.05 to claim evidence for emission at a specific energy. We carried out the same test with events of energies greater than 25 GeV. Figure~\ref{fig5:Fig_SOPIE} (bottom panel) shows the distribution of 58 (17) events above 10 (25) GeV, leading to a p-value of 1e-4 (0.01), thus showing evidence for emission above 10 and, marginally, above 25 GeV. We also tested for possible emission above 30 GeV but found that, despite the presence of 10 events above this energy, their distribution in phase yielded a p-value that was not significant (p $>$ 0.05). 

%\FloatBarrier
\subsection{MAGIC Results} \label{MAGICresults}

\begin{figure}[ht!]
       \centering
         \includegraphics[width=0.6\columnwidth]{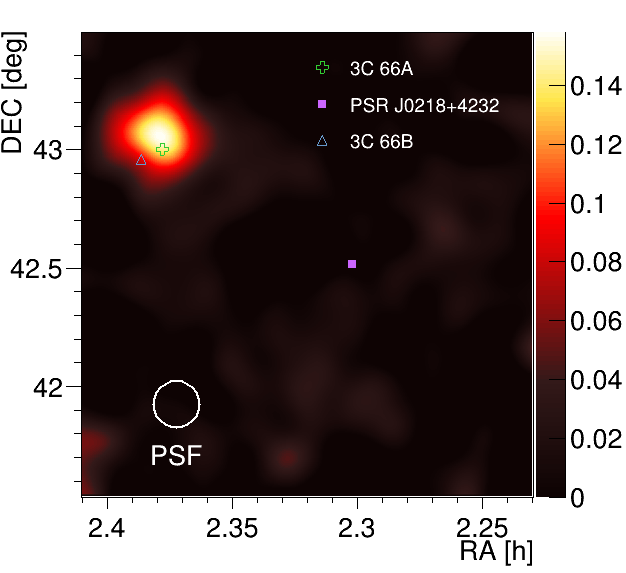}
\caption{MAGIC skymap of the region around PSR J0218+4232 (indicated by a purple square) above 20 GeV.  The relative flux (in arbitrary units) is calculated by the number of smeared excess events divided by the residual background flux within 0.1 degrees \citep{Zanin2013}). Although no VHE emission is detected from J0218, the blazar 3C 66A (green cross), a well-known VHE source~\citep{Acciari09,Aliu09}, is seen with high significance. }
\label{fig7:MAGICskymap}
\end{figure}

\noindent We analyzed our MAGIC data to search for possible pulsed and un-pulsed $\gamma$-ray emission above 20 GeV. The skymap is shown in Figure~\ref{fig7:MAGICskymap}, where no emission is observed from J0218. The high emission spot observed in the image is the blazar 3C 66A, which is significantly detected as a by-product of the observations centered on J0218.

%%\noindent We analyzed our MAGIC data to search for possible pulsed and un-pulsed $\gamma$-ray emission above 20 GeV. The relative fluxes (excess/background) observed in the field of view (FoV) are shown in Figure~\ref{fig7:MAGICskymap}. No emission is observed from J2018. The high emission spot observed in the image is the blazar 3C 66A, which is significantly detected as a by-product of the observations centered on J2018.

\noindent We used the \textsc{Tempo2} package \citep{Hobbs2006} to assign the rotational phase to each event using the same ephemeris as in our LAT analysis described in Section~\ref{LATresults} given in Table~\ref{t:timing}. 

Given the broad pulse shape in the high-energy band, the use of the off-pulse region to estimate the background would lead to large uncertainties due to its smaller phase extent of 0.36 compared to the on-pulse width of 0.64. Therefore, we chose three source-free reflected-region backgrounds located at the same distance from the FoV center, which we expect to have the same acceptance as the region containing the source \citep{Berge2007}. The upper limits (ULs) to the differential flux were obtained by following the \cite{Rolke2001} method under the assumption of a Gaussian systematic uncertainty in the detection efficiency, with a standard deviation of 30\% systematic uncertainty in the flux level. Hereafter, the ULs will be given at 95\% confidence level (CL). We assumed a spectral index $\Gamma$=-4.5 obtained from the power-law fit to the high-energy ($>$10 GeV) \textit{Fermi}-LAT data, as reported in the Third Hard Source Catalog~\citep[][3FHL]{Ajello2017}. 

\noindent Figure~\ref{fig4:LATMAGICSED} shows our MAGIC upper limits, indicated with blue arrows, along with the green points and upper limits from the \textit{Fermi}-LAT analysis. The numerical values are reported in Table \ref{t:LAT_MAGIC_sed_table}.

\begin{table*}
	\caption{ \textit{Fermi}-LAT and MAGIC spectral points and Upper Limits. Centers of energy bins are reported. {\it Fermi}-LAT data utilizes 32 logarithmically spaced bins between 100 MeV and 870 GeV. Three bins, spanning 12.38 - 28.99 GeV, were combined in order to produce a flux point instead of an upper limit. As such, a total of 30 bins are reported for the {\it Fermi}-LAT. MAGIC utilizes 14 logarithmically spaced bins between 20 GeV and 63 TeV. Note that we did not obtain upper limits for the last five MAGIC bins (i.e. E$>3.56$ TeV) because they have zero counts and such limits would be considered too unreliable.} 
	\centering
	\begin{tabular}{lcc|lcc}
	\hline\hline
    \multicolumn{1}{c}{} & \multicolumn{1}{c}{\textbf{Fermi-LAT}} & \multicolumn{1}{c}{\textbf{MAGIC}} & 
    \multicolumn{1}{c}{} & \multicolumn{1}{c}{\textbf{Fermi-LAT}} & \multicolumn{1}{c}{\textbf{MAGIC}} \\
	\multicolumn{1}{l}{}{\textbf{E}} & \multicolumn{1}{c}{}{\textbf{E$^{2}$dN/(dEdAdt})} & \multicolumn{1}{c}{}{\textbf{E$^{2}$dN/(dEdAdt})} & \multicolumn{1}{l}{}{\textbf{E}} & \multicolumn{1}{c}{}{\textbf{E$^{2}$dN/(dEdAdt})} & \multicolumn{1}{c}{}{\textbf{E$^{2}$dN/(dEdAdt})}\\ %\multicolumn{2}{c}{\textbf{E$^{2}$dN/(dE dA dt})} \\
	%	\colhead{\textbf{E$^{2}$dN/(dE dA dt})} & \colhead{} \\
	\multicolumn{1}{l}{}{[GeV]} & \multicolumn{1}{c}{}{[TeV cm$^{-2}$ s$^{-1}$]} & \multicolumn{1}{c}{}{[TeV cm$^{-2}$ s$^{-1}$]} & \multicolumn{1}{l}{}{[GeV]} & \multicolumn{1}{c}{}{[TeV cm$^{-2}$ s$^{-1}$]} & \multicolumn{1}{c}{}{[TeV cm$^{-2}$ s$^{-1}$]} \\
    \hline
	0.12 & (7.24 $\pm$ 1.58) $\times$ 10$^{-12}$ & \nodata & 33.40 & $<$~3.79 $\times$ 10$^{-13}$ & \nodata \\
	0.15 & (7.69 $\pm$ 1.02) $\times$ 10$^{-12}$ & \nodata & 44.35 & $<$~4.71 $\times$ 10$^{-13}$ & \nodata\\ 
	0.20 & (8.11 $\pm$ 0.73) $\times$ 10$^{-12}$ & \nodata & 45.93 & \nodata & $<$~8.32 $\times$ 10$^{-12}$\\ 
	0.27 & (7.77 $\pm$ 0.57) $\times$ 10$^{-12}$ & \nodata & 58.88 & $<$~5.82 $\times$ 10$^{-13}$ & \nodata\\
	0.36 & (8.25 $\pm$ 0.48) $\times$ 10$^{-12}$ & \nodata & 78.18 & $<$~7.69 $\times$ 10$^{-13}$ & \nodata \\ 
	0.48 & (8.15 $\pm$ 0.42) $\times$ 10$^{-12}$ & \nodata & 81.68 & \nodata & $<$~2.36 $\times$ 10$^{-12}$ \\ 
	0.63 & (8.34 $\pm$ 0.38) $\times$ 10$^{-12}$ & \nodata & 103.80 & $<$~1.33 $\times$ 10$^{-12}$ & \nodata \\
	0.84 & (7.50 $\pm$ 0.36) $\times$ 10$^{-12}$ & \nodata & 137.80 & $<$~1.38 $\times$ 10$^{-12}$ & \nodata \\
    1.11 & (8.22 $\pm$ 0.37) $\times$ 10$^{-12}$ & \nodata & 145.25 & \nodata & $<$~6.57 $\times$ 10$^{-13}$ \\
    1.48 & (6.79 $\pm$ 0.36) $\times$ 10$^{-12}$ & \nodata & 183.00 & $<$~1.80 $\times$ 10$^{-12}$ & \nodata \\
    1.96 & (5.34 $\pm$ 0.35) $\times$ 10$^{-12}$ & \nodata & 243.00 & $<$~2.40 $\times$ 10$^{-12}$ & \nodata \\
    2.61 & (5.37 $\pm$ 0.38) $\times$ 10$^{-12}$ & \nodata & 258.30 & \nodata & $<$~3.50 $\times$ 10$^{-13}$\\
    3.46 & (4.05 $\pm$ 0.38) $\times$ 10$^{-12}$ & \nodata & 322.60 & $<$~3.19 $\times$ 10$^{-12}$ & \nodata  \\
    4.59 & (3.07 $\pm$ 0.37) $\times$ 10$^{-12}$ & \nodata & 428.30 & $<$~4.29 $\times$ 10$^{-12}$ & \nodata \\
    6.10 & (2.90 $\pm$ 0.40) $\times$ 10$^{-12}$ & \nodata & 459.34 & \nodata & $<$~1.94 $\times$ 10$^{-13}$ \\
    8.10 & (1.28 $\pm$ 0.30) $\times$ 10$^{-12}$ & \nodata & 568.70 & $<$~5.83 $\times$ 10$^{-12}$ & \nodata  \\
    10.75 & (9.24 $\pm$ 3.01) $\times$ 10$^{-13}$ & \nodata & 755.00 & $<$~8.10 $\times$ 10$^{-12}$ & \nodata \\
    18.95 & (4.66 $\pm$ 1.71) $\times$ 10$^{-13}$ & \nodata & 816.84 & \nodata & $<$~3.30 $\times$ 10$^{-13}$\\
    25.83 & \nodata & $<$~9.12 $\times$ 10$^{-11}$          & 1452.58 & \nodata & $<$~1.06 $\times$ 10$^{-13}$  \\
          &                                       &         & 2583.09 & \nodata & $<$~1.05 $\times$ 10$^{-13}$   \\
\hline
    \end{tabular}
    \label{t:LAT_MAGIC_sed_table}
\end{table*}

\begin{figure}[ht!]
       \centering
                \includegraphics[width=\columnwidth]{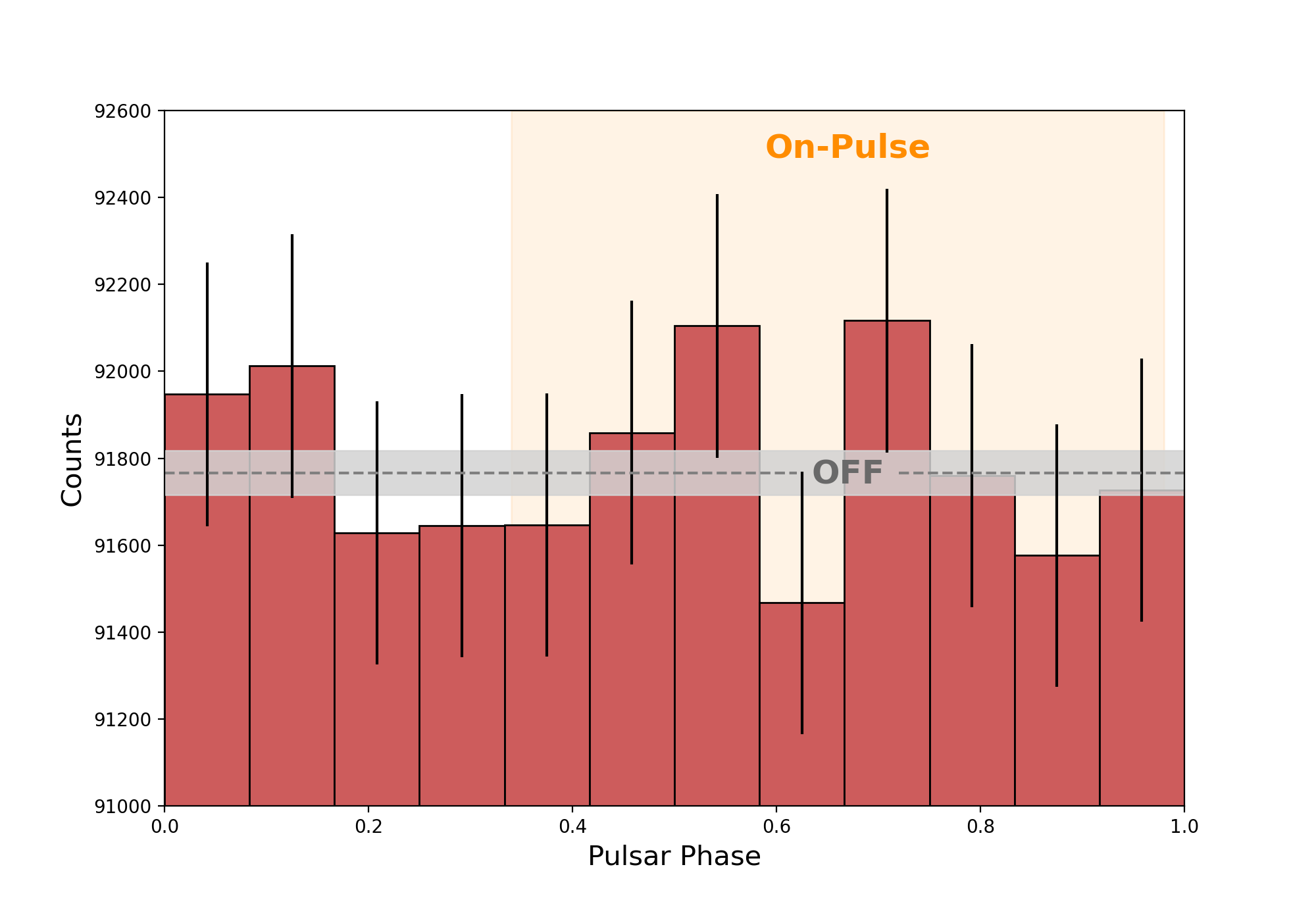}
\caption{Search for VHE pulsations using MAGIC events between 20 - 200 GeV, shown with the pink histogram. We used the same on-pulse interval as LAT analysis, [0.34-0.98], presented with the gold area. The grey horizontal dashed line within the one sigma uncertainty band indicates the average number of OFF events collected from three reflected-region backgrounds in the FoV. No significant pulsation is detected.}
\label{fig9:MAGICphaseo}
\end{figure}

\noindent Given the soft spectrum observed by {\it Fermi}-LAT, we concentrated our search for pulsed signal in the lowest energy decade starting at threshold energy for the MAGIC observations. Hence, the phase-folded light curve of J0218 (see Figure~\ref{fig9:MAGICphaseo}) is computed in the energy range from 20 GeV to 200 GeV. We performed the same unbinned likelihood test described in Section~\ref{LATresults}, to determine whether they are likely to come from the distribution function represented by the lower energy template (see Figure~\ref{fig5:Fig_SOPIE}, top panel), obtaining no evidence for pulsation (p-value $\gg$ 0.05). In addition, we chose to carry out a standard pulsation search, looking at {\it ON} and {\it OFF} events. The on-pulse region was selected as the phase interval between 0.34-0.98 and shown as the gold area in Figure~\ref{fig9:MAGICphaseo}, as defined by our \textit{Fermi}-LAT analysis (see the top panel of Figure~\ref{fig5:Fig_SOPIE} and Table~\ref{t:timing}). The same source-free reflected-region backgrounds, shown with the grey horizontal band in Figure~\ref{fig9:MAGICphaseo}, were used for calculating the significance of the excess events using Eq.17 of \citep{LiMa1983}), and no significant ($0.057\sigma$) pulsation was found. 
%by Li \& Ma formula( Eq.17 from \citep{LiMa1983}). No significant pulsation was found (see Table~\ref{t:MAGIC_phaseostats_table}).
Moreover, we applied region-independent signal tests \citep{deJager1989} ( $\chi^{2}$, and H-test), also with null results (5.54, for 11 degrees of freedom, and $0.05\sigma$, respectively).

%\begin{deluxetable}{ll}
%	\tablecaption{\label{t:MAGIC_phaseostats_table} MAGIC Pulsed emission Signal Tests' Result}
%	\tablecolumns{2}
%	\tablewidth{0pt}
%	\tablehead{ \colhead{Tests} & \colhead{Results}\\ }
%	\startdata
%	Li\&Ma & $5.7 \times 10^{-2} \sigma$ \\
%	$Z_{10}^{2}$ & $2.0 \times 10^{-1} \sigma$\\
%	H-test & $5.0 \times 10^{-2} \sigma$\\ 
%\enddata	
%\end{deluxetable}

\FloatBarrier
\section{Theoretical Modeling} \label{theory}

\noindent We modeled the broadband spectrum of J0218 from UV to VHE $\gamma$-rays (14 orders of magnitude in energy) using a numerical force-free magnetosphere model for the global magnetic field, computing the individual trajectories of particles injected at the neutron star surface. Two populations of particles are injected: primary electrons/positrons along field lines that connect to the current sheet and are accelerated by an assumed parallel electric field distribution, and secondary electrons/positrons from polar cap pair cascades along field lines where there is no accelerating electric field. The dynamics and radiation of the particles are followed from the neutron star surface to a distance of 2 light cylinder radii (2$R_{lc}$) and radiated photons are stored in energy-dependent sky maps of observer angle vs. rotation phase~\citep{Harding15,Harding18}.

\noindent All particles radiate by synchro-curvature (SC) and inverse Compton (IC) emission. The pitch angles for SC are maintained through cyclotron resonant absorption of radio photons emitted above the polar cap (PC)~\citep{Harding08}. The SC also assumes the radius of curvature of the particle trajectory in the inertial observers' frame. The IC requires that trajectories be followed twice, once to store the SC radiation emissivity and another to compute the local photon densities from the stored emissivity and radiate IC~\citep{Harding15,Harding18}.  

\begin{figure}[ht!]
\centering
\includegraphics[width=6.5in]{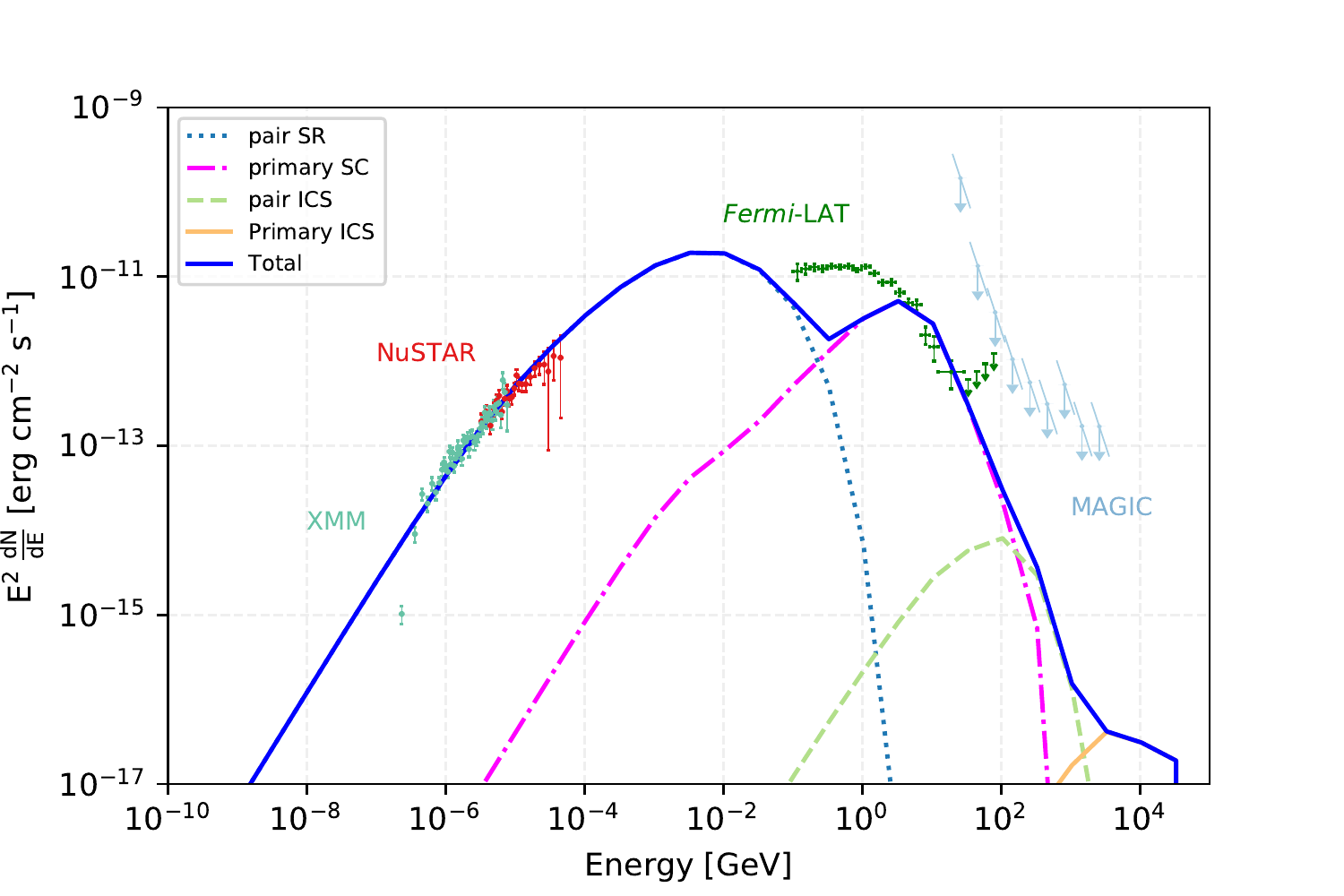} 
\caption{Model predictions for the spectrum of phase-averaged emission from accelerated particles and pairs in PSR J0218+4232, for an assumed magnetic inclination angle $\alpha$ = 45$\arcdeg$ and viewing angle $\zeta$ = 65$\arcdeg$. The solid orange line represents the predicted ICS component due to accelerated SC-emitting primaries scattering the pair SR component (blue dotted line). The thick blue line identifies the overall emission model. Data points show the soft (XMM) and hard (NuSTAR) X-ray emission~\citep[from][]{Gotthelf2017}, as well as the LAT spectral points and MAGIC upper limits obtained in this work. Note that the LAT and MAGIC spectral points and upper limits represent the total (pulsed plus unpulsed) emission, however, given the broad peak and large pulsed fraction (see Figure~\ref{fig5:Fig_SOPIE}), the differences between this and the {\it pulsed} spectrum would be marginal, and would not significantly affect the model fit parameters. Note that the $\gamma$-ray data are identical to those in Figure~\ref{fig4:LATMAGICSED}. }
\label{fig10:TheorySSC}
\end{figure}

\noindent The main assumptions of the model are the parallel electric field ($E_{||}$) distribution, the source of pairs, pair multiplicity and their injection distribution on the PC, and the mechanism for generating pitch angle. Apart from this, the model requires the observed parameters of the pulsar ($P$ and $\dot{P}$). The magnetic and electric field distribution assumptions are based on results of Particle-In-Cell (PIC) simulations showing that pulsars producing high pair multiplicity have near-force-free magnetospheres and that the highest parallel electric fields are in the current sheet. Increasing the pair multiplicity increases the pair SC (mostly synchrotron radiation (SR)) and the pair IC (mostly synchrotron self-Compton, SSC). Increasing the E parallel increases the SC of primaries, increases the high-energy (GeV) cutoff, and increases the IC (at 10 TeV). For J0218, we assumed a magnetic inclination angle $\alpha$ = 45$\arcdeg$, a viewing angle $\zeta$ = 65$\arcdeg$, and a pair multiplicity of M$_+$ = 1$\times10^5$. Figure~\ref{fig10:TheorySSC} shows the model predictions, including the various individual emission components. 

\vspace{1cm}

\noindent We have also used a synchro-curvature model where all unknowns are reduced to just a few parameters that represent the observed spectrum ~\citep[see][for details]{Torres18,Torres19}. The model follows particle trajectories in a generic region of a pulsar magnetosphere threaded by an accelerating parallel electric field, $(E_{||})$. The region is located around the light cylinder, and particles are assumed to enter it at $x_{in}$ with a (sizeable) pitch angle $\alpha$. The model parameterizes the magnetic field by a power law $B(x) = B_s (R_s/x)^{b}$ (see the discussion in \cite{Vigano2015-2}), where $x$ is the distance along the field line, $b$ is referred to as the magnetic gradient, $B_s$ is the surface magnetic field, and $R_s$ is the pulsar radius. Given $(E_{||}, b)$ as free parameters, and the period and period derivative $(P,\dot P)$, the model solves the equations of motion that balance acceleration and losses by SC radiation (see \cite{Cheng96,Vigano2015-1}), computing the total emission. The model assumes that the distribution of particles emitting towards us can be parameterized as ${dN_e}/{dx} \propto  e^{-(x-x_{\rm in})/x_0} $ where the inverse of $x_0/R_{lc}$ ($R_{lc}$ is the light cylinder radius) is referred to as the {\it contrast}. 

\begin{figure}[ht!]
\begin{center}
\includegraphics[width=0.9\textwidth]{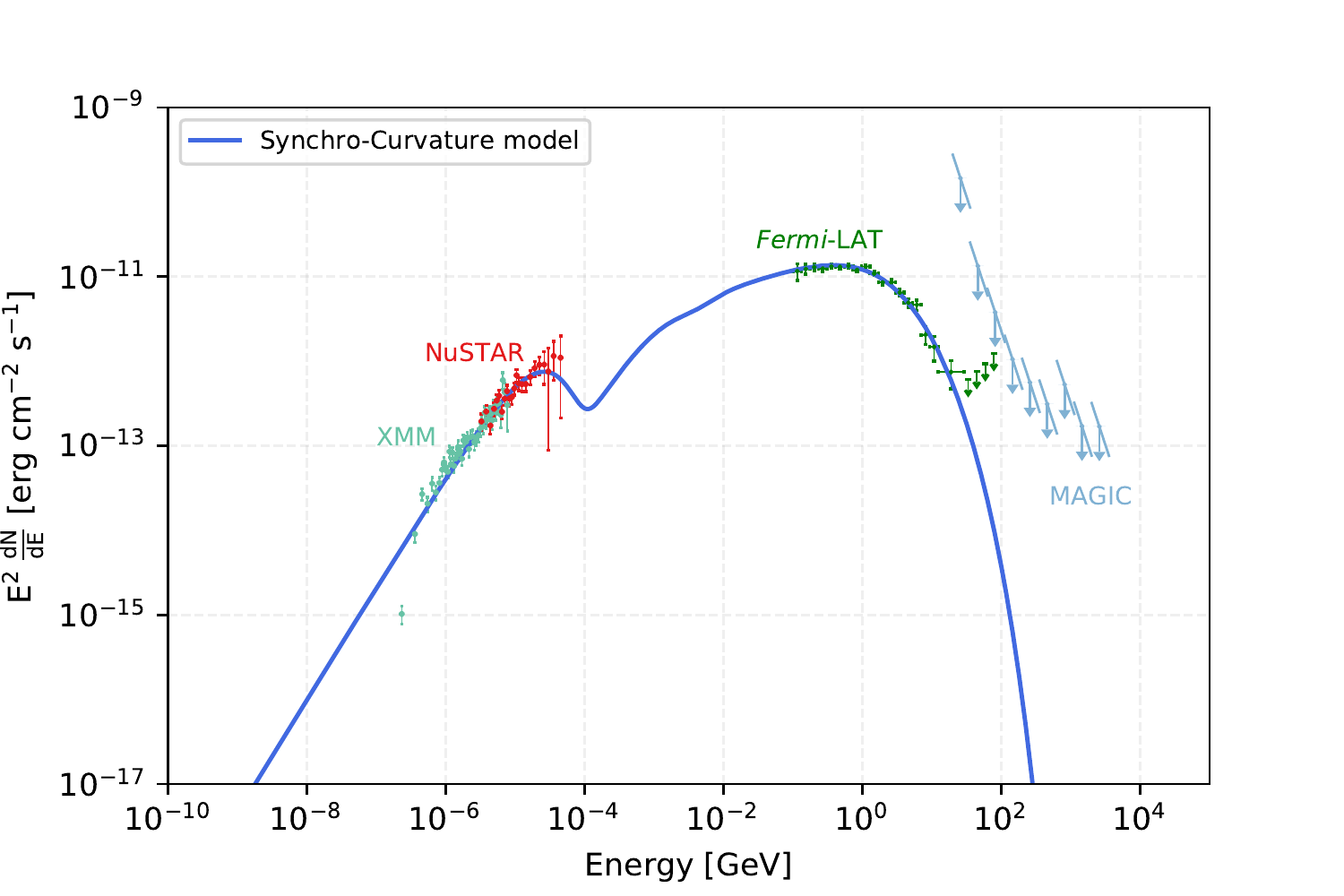}
\vspace{-0cm}
\caption{Broad-band spectrum of PSR J0218+4232, from the X-ray (XMM and NuSTAR) to the $\gamma$-ray ({\it Fermi}-LAT and MAGIC) range, along with the best fit to the synchro-curvature model~\citep{Torres18}. 
The model is described by these parameters:
$log(E_{||}/V\,m^{-1})$=10.92, 
%$logE_{||}$=10.92, 
$\log(x_0/R_{lc})$=$-$4.20,
%$log(\frac{x_{0}}{R_{lc}})$=$-$4.20, 
and $b$=3.70. 
Note that the X-ray and $\gamma$-ray data are identical to those in Figure~\ref{fig10:TheorySSC}.
%\replaced{Note that the X-ray and $\gamma$-ray data are identical as in Figure~\ref{fig10:TheorySSC}.}{Note that the X-ray and $\gamma$-ray data are identical to those in Figure~\ref{fig4:LATMAGICSED}.}
}
\label{fig11:TheoryTorres}
\end{center}
\end{figure}

\noindent Figure \ref{fig11:TheoryTorres} shows the results of the model with best fit parameters, $log(E_{||}/V\,m^{-1})$=10.92, 
%$logE_{||}$=10.92 with $E_{||}$ in units of V/m, 
$\log(x_0/R_{lc})$=$-$4.20 and $b$=3.70. 
The agreement between the model description and the broad-band data is acceptable (the fractional residual errors are of the order $\sim$10\%), despite the significant increase in the precision of each spectral measurement and the number of data points.

\section{Discussion and Conclusions} \label{conclusion}

\noindent The detection by ground-based Cherenkov telescopes of pulsed emission from the Crab pulsar~\citep{Aliu08,VERITAS_Crab}, most recently detected up to TeV energies~\citep{Ansoldi16}, followed by the detection of pulsations from Vela~\citep{HESS_Vela}, also up to TeV energies~\citep{Harding18b}, has led to a flurry of activity by pulsar experts to develop self-consistent models able to explain the detected emission over such a broad range of energies.

\noindent In addition, the development of the Sum-Trigger-II system in MAGIC has significantly improved the sensitivity of the telescopes below 100 GeV, something which has enabled the detection of Geminga between 15 GeV and 75 GeV~\citep{Acciari20}, making this the third $\gamma$-ray pulsar (and first {\it middle-aged} one) detected with ground-based telescopes~\footnote{The H.E.S.S. Collaboration has also reported at the 36th International Cosmic Ray Conference (ICRC2019) the detection of $\gamma$-ray pulsations up to $\sim$70 GeV from PSR B1706--44~\citep{Spir-Jacob19}, which, if confirmed, would bring the total number of ground-based detected $\gamma$-ray pulsars to four.}. 

\noindent J0218 is part of a small but diverse population of MSPs with a well-characterized broad-band non-thermal energy distribution. Several of these MSPs have an `inverted’ spectrum in X-rays (where $E^2 dF/dE$ decreases with energy), quite different from that of J0218 as we have reported here, see Figure 3 of \cite{Zelati2020}. 
\noindent We also note that the fitted magnetic gradient $b$ for J0218 and other MSPs within the synchro-curvature model~\citep{Torres18} is larger than for normal pulsars. This is perhaps the result of the larger $B_{lc}$ of MSPs compared to typical pulsars, due to the smaller size of $R_{lc}$. This needs to be taken into account when making predictions for their observability at lower energies based only on the $\gamma$-ray data. Fits to the $\gamma$-ray data alone are mostly insensitive to the value of the magnetic gradient, and assuming a lower $b$ could lead to incorrect predictions that an MSP is undetectable in the X-ray band.

\noindent In accordance with previous studies, we also find here that the relevant scales for the production of the pulsar's spectrum (given by $x_0$) is small in comparison with the light cylinder radius.

\noindent This is true in general for MSPs, for which the light cylinder is already orders of magnitude smaller than in normal pulsars, i.e., the $x_0/R_{lc}$-values imply a relevant region of emission $\ll 1$ km.

\noindent Instead of subtracting the background events from the off-pulse region, we applied reflected-region background subtraction approach for the MAGIC analysis due to the large on-pulse interval of the LAT phaseogram. No evidence of emission (either pulsed or unpulsed) is apparent in the MAGIC data, and the measured MAGIC upper limits are well above our two theoretical model predictions for VHE emission. The curvature radiation component from particles accelerated mostly in the current sheet is expected to fall to flux levels too low at VHE energies for detection by MAGIC and the ICS components from both pairs (mostly SSC) and accelerated primaries are predicted to be at even lower flux levels.

\noindent Most models for $\gamma$-ray emission from pulsars do not predict high levels of ICS and SSC emission for MSPs. In the model we used here, described by \cite{Harding15} and \cite{Harding18}, for example, the pairs that come from the PC cascade and MSP surface magnetic fields are so low that the photons need to have much higher energies to produce pairs by one-photon magnetic pair production than do normal pulsars. The MSP pair spectra are thus shifted to much higher energies (typically $\gamma \sim 10^4 - 10^7$) \citep{Harding11}. This will produce higher energy SR near the light cylinder. Since VHE emission is most likely ICS or SSC, and both particles and photons have higher energies, the VHE emission will be Klein-Nishina limited and therefore suppressed. This is also a problem for outer gap (OG) models since the latest models have pairs also produced near the PC since otherwise, MSPs cannot sustain OGs~\citep{Harding21}. Observationally, we see that the SR spectra seem to extend to higher energy in MSPs (at least the energetic ones that have non-thermal emission).  So the SR photons and the particles that produce them must be at higher energy. The Cherenkov Telescope Array (CTA) is expected to have significantly better sensitivity than MAGIC in the 10--100 GeV range, and this and other pulsars will thus be prime targets for observation~\citep{burtovoi17}. On the other hand, pulsars like J0218 are also good sources for MeV telescopes, such as AMEGO~\citep{AMEGO}, that can detect the predicted SR peaks around 1 - 10 MeV.

%\section{Acknowledgments}
\begin{acknowledgments}

% This is the standard Fermi acknowledgement note

\noindent The \textit{Fermi}-LAT Collaboration acknowledges generous ongoing support from a number of agencies and institutes that have supported both the
development and the operation of the LAT as well as scientific data
analysis. These include the National Aeronautics and Space
Administration and the Department of Energy in the United States, the
Commissariat \`{a} l'Energie Atomique and the Centre National de la
Recherche Scientifique / Institut National de Physique Nucl\'{e}aire et
de Physique des Particules in France, the Agenzia Spaziale Italiana
and the Istituto Nazionale di Fisica Nucleare in Italy, the Ministry
of Education, Culture, Sports, Science and Technology (MEXT), High
Energy Accelerator Research Organization (KEK) and Japan Aerospace
Exploration Agency (JAXA) in Japan, and the K. A. Wallenberg
Foundation, the Swedish Research Council and the Swedish National
Space Board in Sweden. 
This research was partially carried out using the HKU Information Technology Services research computing facilities that are supported in part by the Hong Kong UGC Special Equipment Grant (SEG HKU09). P.S.P. and C.Y.N. are supported at HKU by a grant from the Big Data Project Fund (BDPF) and a GRF grant (Project 17307618) from the Hong Kong Government. AKH acknowledges resources supporting the theoretical modeling provided by the NASA High-End Computing Program through  the NASA Center for Climate Simulation. DFT acknowledges support from the Spanish grants PGC2018-095512-B-I00, SGR2017-1383; as well as from the Chinese Academy of Sciences Presidential Fellowship Initiative 2021VMA0001. The Nan\c{c}ay Radio Observatory is operated by the Paris Observatory, associated with the French Centre National de la Recherche Scientifique (CNRS). We acknowledge financial support from the ``Programme National Hautes Energies'' (PNHE) of CNRS/INSU, France.  \\
\\
% This is the standard MAGIC acknowledgement note
\noindent We would like to thank the Instituto de Astrof\'{\i}sica de Canarias for the excellent working conditions at the Observatorio del Roque de los Muchachos in La Palma. The financial support of the German BMBF, MPG and HGF; the Italian INFN and INAF; the Swiss National Fund SNF; the ERDF under the Spanish Ministerio de Ciencia e Innovaci\'{o}n (MICINN) (FPA2017-87859-P, FPA2017-85668-P, FPA2017-82729-C6-5-R, FPA2017-90566-REDC, PID2019-104114RB-C31, PID2019-104114RB-C32, PID2019-105510GB-C31,PID2019-107847RB-C41, PID2019-107847RB-C42, PID2019-107988GB-C22); the Indian Department of Atomic Energy; the Japanese ICRR, the University of Tokyo, JSPS, and MEXT; the Bulgarian Ministry of Education and Science, National RI Roadmap Project DO1-268/16.12.2019 and the Academy of Finland grant nr. 320045 is gratefully acknowledged. This work was also supported by the Spanish Centro de Excelencia ``Severo Ochoa'' SEV-2016-0588 and CEX2019-000920-S, and ``Mar\'{\i}a de Maeztu'' CEX2019-000918-M, the Unidad de Excelencia ``Mar\'{\i}a de Maeztu'' MDM-2015-0509-18-2 and the ``la Caixa'' Foundation (fellowship LCF/BQ/PI18/11630012) and by the CERCA program of the Generalitat de Catalunya; by the Croatian Science Foundation (HrZZ) Project IP-2016-06-9782 and the University of Rijeka Project 13.12.1.3.02; by the DFG Collaborative Research Centers SFB823/C4 and SFB876/C3; the Polish National Research Centre grant UMO-2016/22/M/ST9/00382; and by the Brazilian MCTIC, CNPq and FAPERJ. We thank Matthew Kerr and Philippe Bruel for their careful reading of the manuscript and useful comments and suggestions, which greatly improved the paper. We thank Slavko Bogdanov for providing the X-ray data points used in this work. Finally, we sincerely thank the journal referee for taking the time and effort to provide a detailed review of our paper, providing some very helpful feedback, which helped us improve the final version of the article.

\end{acknowledgments}

\vspace{5mm}

\software{Fermipy~\citep{wood17}, Minuit~\citep{James1994}, MARS~\citep{Zanin2013}, R~\citep{R13}, SOPIE~\citep{SOPIE}, Tempo2~\citep{Hobbs2006}}

\facilities{\textit{Fermi}-LAT, MAGIC, ADS}

\vspace{3mm}

\bibliography{ms}{}
\bibliographystyle{aasjournal}



\end{document}